\documentclass[conference]{IEEEtran}
\IEEEoverridecommandlockouts
\usepackage{cite}
\usepackage{amsmath,amssymb,amsfonts}
\usepackage{algorithmic}
\usepackage[table]{xcolor}
\usepackage{graphicx}
\usepackage{textcomp}
\usepackage{xcolor}
\usepackage{balance}
\usepackage{hyperref}
\usepackage{multirow}
\usepackage{multicol}
\usepackage{tabularx}
\usepackage{xspace}
\usepackage{comment}

\newcommand{\ie}{\textit{i.e.}\xspace}
\newcommand{\eg}{\textit{e.g.}\xspace}

\def\BibTeX{{\rm B\kern-.05em{\sc i\kern-.025em b}\kern-.08em
    T\kern-.1667em\lower.7ex\hbox{E}\kern-.125emX}}
\begin{document}

\title{Predicting Code Review Completion Time in Modern Code Review}

\author{
\IEEEauthorblockN{
Moataz Chouchen\IEEEauthorrefmark{1},
Jefferson Olongo\IEEEauthorrefmark{1},
Ali Ouni\IEEEauthorrefmark{1},
Mohamed Wiem Mkaouer\IEEEauthorrefmark{2}}
\IEEEauthorblockA{\IEEEauthorrefmark{1}ETS Montreal, University of Quebec, QC, Canada}
\IEEEauthorblockA{\IEEEauthorrefmark{2}Rochester Institute of Technology, Rochester, NY, USA}
\{moataz.chouchen.1,aurelien-jefferson.olongo-onana-noah.1\}@ens.etsmtl.ca, ali.ouni@etsmtl.ca, mwmvse@rit.edu 
}

\maketitle

\begin{abstract}
\textbf{Context.} 
Modern Code Review (MCR) is being adopted in both open source and commercial projects as a common practice. MCR is a widely acknowledged quality assurance practice that allows early detection of defects as well as poor coding practices. It also brings several other benefits such as knowledge sharing, team awareness, and collaboration. %To maintain high code review quality, it is crucial to effectively manage the review process. %estimate the effort required for a given code review since effort estimation help developers to better prioritize review requests.

\textbf{Problem.} In practice, code reviews can experience significant delays to be completed due to various socio-technical factors which can affect the project quality and cost. For a successful review process, peer reviewers should perform their review tasks in a timely manner while providing relevant feedback about the code change being reviewed. However, there is a lack of tool support to help developers estimating the time required to complete a code review prior to accepting or declining a review request. %Effectively predicting the code review completion time allows developers to estimate their effort needed for a given review request and better manage and prioritize their reviews.%ii) Alert developers about code reviews taking more than expecting completion time. \ali{but, what is the problem?} 
 %While MCR research is becoming more mature, predicting the effort for code reviews in the context MCR is not addressed.
%\ali{hmm.. this looks weak. "It is not addressed" could be because it is not important, so no one care about it. There are many problems that are not addressed, and better to keep them not addressed, as addressing them could be just a waist of time or might bring little/negligible impact in practice. Rather, explain why it is important to predict completion time; and what happens if we don't predict it. Take home message: in research never say "no one address it so I am going to address it"}. 
%In fact, the studies focusing on prioritizing code reviews in MCR did not provide an approach to quantify the effort needed for a code review. 

\textbf{Objective.} Our objective is to build and validate an effective approach to predict the code review completion time in the context of MCR and help developers better manage and prioritize their code review tasks.

\textbf{Method.} We formulate the prediction of the code review completion time as a learning problem. In particular, we propose a framework based on regression models to (\textit{i}) effectively estimate the code review completion time, and (\textit{ii}) understand the main factors influencing code review completion time. 
\end{abstract}


\section{Introduction}

%\ali{the flow of the introduction needs rework. This is a typical structure of the introduction}

%\textbf{Paragraph 1: General paragraph\\}
Code review (originally known as code inspection) is a crucial activity in software development where team members review, provide comments, and criticize code changes prior to merging a code change in the main codebase \cite{fagan2002design,ackerman1984software}. Since traditional code review processes were cumbersome and time-consuming, code review practices have migrated to modern code review (MCR), a lightweight, tool-based and asynchronous process \cite{beller2014modern}. Since its emergence, MCR has been widely adopted, and supported with collaborative cloud-based tools in both open-source and commercial software projects such as \textit{Gerrit}\footnote{\url{https://www.gerritcodereview.com}} and \textit{ReviewBoard}\footnote{\url{https://www.reviewboard.org}}. MCR has show efficiency in helping developers to improve the quality of their code and reduce post-integration defects \cite{macleod2017code,bacchelli2013expectations,sadowski2018modern}. 

%\textbf{Paragraph 2: Problem statement (precise definition + importance)\\}

While MCR has its own advantages, it also presents several challenges in practice. For instance, Patanamon et al. \cite{Patanamon2015revfinder} have shown that code reviews can take an average of 14 days to be reviewed. To help managing the code review process and reduce the completion time, several approaches have been proposed to recommend appropriate reviewers for pending code reviews \cite{ouni2016search,kovalenko2018does,hannebauer2016automatically,Balachandran2013icse,Patanamon2015revfinder}. However, most of these approaches tend to recommend reviewers who are most experienced with the code change without considering the required time to complete the code review. Other studies proposed approaches to prioritize code review requests \cite{fan2018early,saini2021using} in the context of MCR. However, still, the estimated completion time is not considered to better prioritize code reviews. %Additionally, to our best knowledge, only Maddila et al. \cite{maddila2019predicting} attempted to estimate pull request effort. However no studies attempted to quantify the effort of a code review request in MCR. 

In a typical software project, developers can receive several review requests daily.
To better manage and prioritize their code review activities, developers need to estimate the review time for their review requests which can be tedious and error-prone, especially when multiple files are involved in the code change \cite{gousios2015work,maddila2019predicting,chouchen2021anti}. That is, inappropriate estimation of code review completion time can lead to unforeseen delays in the whole code review process and software product delivery \cite{bacchelli2013expectations,chouchen2021whoreview}. Hence, it is crucial to provide efficient and accurate approaches to estimate code review completion time since it will help practitioners to better optimize their code review process.

Prior research has shown that the code review process can be impacted by several socio-technical factors such as the complexity of the code change, the context of the change, the author's experience and collaboration, and reviewers' participation and the communication environment \cite{baysal2016investigating,alomar2021refactoring,chouchen2021anti,hirao2020code,ruangwan2019impact}. Hence, predicting the completion time of a given code review would not be straightforward given that multiple technical and social factors can interfere with it.

%\textbf{Paragraph 3: Existing solution (identify their limitations...)}

%To this end, estimating code review request effort is not yet to be investigated in depth in the context of MCR. In fact, most of the studies focused on proposing approaches to prioritize code review requests \cite{fan2018early,saini2021using} in the context of MCR without providing a way to estimate the effort of a code review request. Additionally, to our best knowledge, only Maddila et al. \cite{maddila2019predicting} attempted to estimate pull request effort. However no studies attempted to quantify the effort of a code review request in MCR. 

%\textbf{Paragraph 4: Contributions (why your solutions are expected to be better, essence of the idea...)\\}
In this paper, we aim at building an automated approach to predict the time required to perform a code review prior to accepting or declining a review request, in the context of MCR. More specifically, we plan to formulate the code review completion time prediction as a learning problem and use Machine Learning (ML) techniques to build prediction models to estimate the code review completion time. Additionally, we plan to exploit the built prediction model in order to investigate the most important factors that impact code review completion time. 
%\textbf{Paragraph 5:  Analysis (theoretical, experimental, simulation...)\\}
%To achieve this, we propose a framework for code review duration prediction that enable us to compare the performance of different ML techniques based on a robust validation process. 
%We anticipate that our framework can be used later as a standard to study code review duration prediction in more depth.
%\textbf{Paragraph 6: results\\}

%\textbf{Paragraph 7: Talk about the paper (registered report) sections\\}
In this registered report, we set up the research protocol that we will follow to set up our framework. The main goals of this study are (\textit{i}) investigate the efficiency of ML techniques to predict the code review completion time, and (\textit{ii}) understand the main factors influencing the code review completion time by investigating the learned models. We anticipate our toot to be integrated into MCR tools as a bot and help reviewers estimating the code review duration prior to accepting or declining a review request.

The remainder of the paper is organized as follows. In section \ref{sec:backround}, we present the necessary background and related work. In section \ref{sec:framework_overview}, we introduce our framework and define its components. Section \ref{sec:empirical_study}is devoted to presenting the details of our experimental study. Section \ref{sec:analysis_method}presents the analysis method for each research question.

%\ali{very early talking about the prediction. You need to make a "story" for your paper. For example: first explain why completing code review in a timely manner is important, and how delayed code reviews can impact the project progress... (Pick have shown in her SANER 2015 paper, that code review take on average 14 days) Then most existing solutions attempted to recommend appropriate reviewers [references] with the goal of reducing the review completion time. However, while this can help, it is still problematic as most of these recommendation approaches tend to recommend expert active reviewers who might be busy with several other open reviews [references]. Solution: introducing a code review completion time prediction technique can help better managing the code review process and reduce its costs... You also need to show some examples from your dataset (with their URL) to showcase these delayed reviews...}

\section{Background and related works}
\label{sec:backround}
\subsection{Modern code review process}

\begin{figure}[h!]
    \centering
    \includegraphics[width=\linewidth]{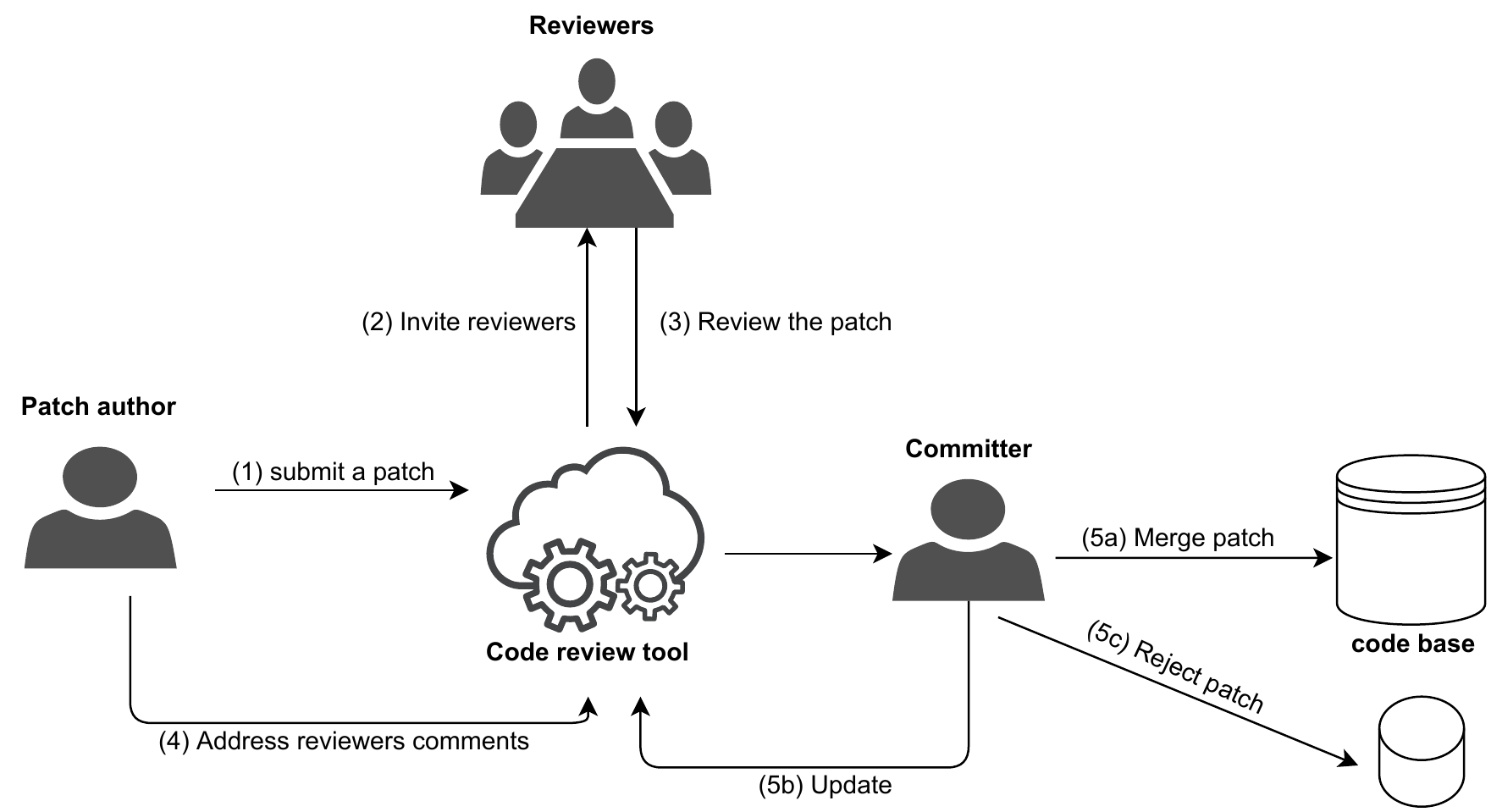}
    \caption{An illustration of MCR process.}
    \label{fig:mcr_process}
\end{figure}

Figure \ref{fig:mcr_process} illustrates the process of MCR supported by \textit{Gerrit} since \textit{Gerrit} is a widely adopted MCR tool. This process starts with a developer submitting a code patch to the code review tool (1). The tool invites reviewers to review the patch (2). The reviewers who accepted the invitation review the patch (3) and the author address reviewers' comments through multiple revisions (4). Finally, a committer who is an experienced developer provides his final decision to: 5a) Merge the patch, 5b) Request more revisions, or 5c) Reject the patch. 
\subsection{Factors influencing code review duration}
Recent works showed that various factors can influence the code review duration. Jiang et al. \cite{jiang2013will} showed the number of reviewers may impact the review time. Bosu et al. \cite{bosu2014impact} showed that core developers receive quicker first feedback on their review requests, complete the review process in a shorter time. Additionally, Baysal et al. \cite{baysal2016investigating} showed through an empirical study on \texttt{Webkit} and \texttt{Google Blink} that technical (\ie, patch size and component) and non-technical (\ie, organization, author experience, and reviewers activity) factors may impact the code review duration. Recently, AlOmar et al. \cite{alomar2021refactoring} showed that, in Xeros, refactoring code reviews take a longer duration than non-refactoring code reviews. 

Previous studies attempted to understand the main factors impacting code review duration. We build upon the results obtained in these studies to extract features characterizing code review completion time in MCR. 

\subsection{Effort estimation in software engineering}
Effort estimation is an ongoing challenge in software engineering research \cite{trendowicz2008state,britto2014effort,sharma2017systematic}. Early attempts of effort estimation in software engineering started by using regression analysis to extract the factors influencing software effort \cite{barry1981software}. Later, a variety of techniques have been proposed for effort estimation based on artificial techniques namely, decision trees \cite{kocaguneli2012active}, support vector machines \cite{oliveira2006estimation} and genetic programming \cite{ferrucci2010genetic,sarro2016multi,tawosi2021multi}. 

In the context of MCR, multiple techniques have been proposed to guide developers to prioritize their code review effort. Fan et al. \cite{fan2018early} proposed an approach to detect whether a given code review is going to be merged helping developers to focus their effort on code reviews that are likely to be merged. Later Saini et al. \cite{saini2021using} proposed a tool for prioritizing code reviews based on Bayesian networks. In the context of pull request development, Maddila et al. \cite{maddila2019predicting} proposed a tool for pull request duration. However, since gerrit-based code review process and pull request code review processes slightly differ in the way how to model a change and gerrit focuses more on formalizing the review process, we plan to study the code review duration prediction problem in the context of gerrit-based code review process.
\section{Framework overview}
\label{sec:framework_overview}
\begin{figure}[!ht]
    \centering
    \includegraphics[scale=0.8]{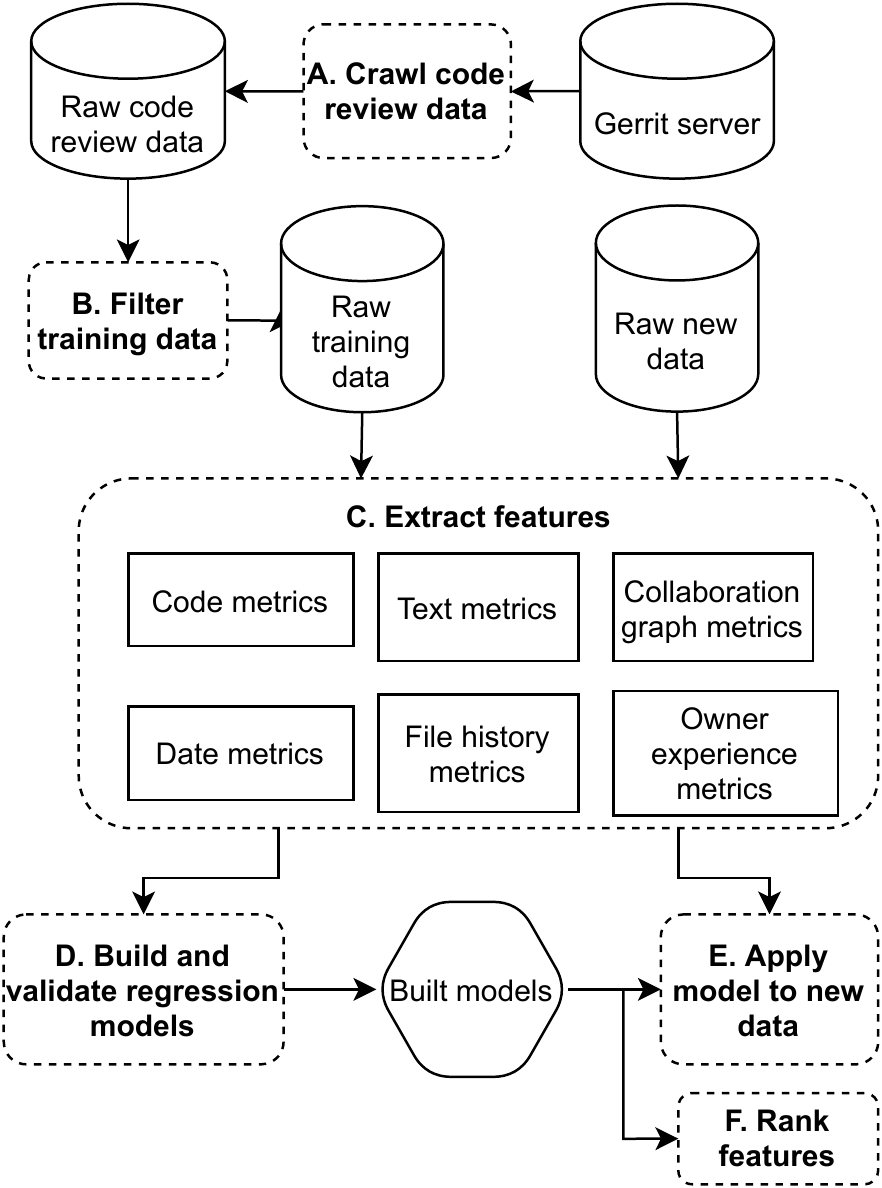}
    \caption{Framework overview.}
    \label{fig:framework_overview}
\end{figure}
In this study, we investigate whether it is possible to formulate code review completion time as a learning problem. Moreover, we investigate the most important factors influencing the completion time of a code review. To achieve these goals, we follow the framework depicted in Figure \ref{fig:framework_overview}. First, we collect raw training data from open source projects that practice MCR, focusing, in particular, on Gerrit-based projects (Step A). Then, we filter our irrelevant data and outliers from the code review data (Step B). 
Thereafter, different features are calculated for the whole training data (Step C). This training data is used later to build and validate our regression models (Step D). Finally, the built regression models will be used either to predict delay for new code reviews (Step E) or to rank features and explore factors influencing code review duration (Step F). In the following, we explain in more detail each step of our framework. 

\subsection{Crawl code review data}
We start by extracting code reviews data and commits for a number of open-source software (OSS) projects from the Gerrit server using \texttt{Gerrit REST API}\footnote{https://gerrit-review.googlesource.com/Documentation/rest-api-changes.html}. Since the first revision commit data is not usually available in the cloned repositories, we used the \texttt{git fetch} command to fetch the first revisions data from the remote servers. 

\subsection{Filter data}
To build robust models with good performance, training data instances, \ie, code reviews, should be carefully selected from the selected projects. In this study, we use the following filers to discard: 
\begin{enumerate}
    \item Re-opened code reviews since these reviews have been already reviewed.
    \item Code reviews that are self-reviewed, \ie, self-merged or self-abandoned by the owner of the code review request since there is no review activity in there. 
    \item Code reviews with a short delay, \ie, with delay $\leq 24$ hours since these reviews can be related to urgent issues (security change or bug with high severity) and should be reviewed at a fast pace \cite{maddila2019predicting}. 
    \item Code reviews with long delays, \ie, more than 3 weeks since these code reviews can be forgotten by the authors or have a low priority. %\ali{First and second quartile}
   % \item \ali{filter out all code reviews that are done only by bots.}
    \end{enumerate}
    
After filtering these irrelevant code reviews, we keep the obtained review data and use it as a training dataset to build our regression models.

\subsection{Extract features}
In our study, we extend the set of features defined by Fan et al. \cite{fan2018early} since it is tightly related to our context. 
In total, we extracted 50 features that are  related to code review completion time prediction. Table \ref{tab:features} summarizes our preliminary list of metrics grouped into six main dimensions: Code features, Text features, Collaboration graph features, Date features, File history features, and Owner experience features. Other metrics can be added to our list according to the obtained results later in this study.

\begin{table*}[ht]
    %\rowcolors{2}{gray!25}{white}
	\fontsize{7}{8.5}\selectfont
	\tabcolsep=0.1cm
     \center
     \caption{Summary of features considered in our studies.}
     %\resizebox{\textwidth}{!}{
     \begin{tabular}{p{2cm}p{6.5cm}p{9cm}}
    %\begin{tabular}{lll}
        \hline
    \textbf{Feature dimension} &\textbf{Feature name}&\textbf{Definition} \\
    %\hline 
    \hline
    \multirow{4}{*}{Date metrics}
     &\texttt{days\_of\_the\_weeks\_of\_date\_created} &The day of the week the change was created \\
     &\texttt{is\_created\_date\_a\_weekend} &Whether the change was created in the weekend\\
     &\texttt{author\_timezone} &The timezone of the author of the change\\
     \hline
    \multirow{6}{*}{\begin{tabular}[C]{l} Collaboration  \\ graph metrics\end{tabular}} &\texttt{degree\_centrality}&\multirow{6}{7.5cm}{Collaboration metrics used to quantify the degree of collaboration of the code review owner in the corresponding project based on prior code review activities as defined in\cite{zanetti2013categorizing,fan2018early}. }\\
     &\texttt{closeness\_centrality} &\\
     &\texttt{betweenness\_centrality} &\\
     &\texttt{eigenvector\_centrality} &\\
     &\texttt{clustering\_coefficient} &\\
     &\texttt{core\_number}&\\
     \hline
    \multirow{9}{*}{Code metrics}&\texttt{\#lines\_added} &Number of lines inserted in the first revision \cite{fan2018early}\\
     &\texttt{\#lines\_deleted} &Number of lines deleted in the first revision \cite{fan2018early}\\
     &\texttt{Code\_churn} &\texttt{\#lines\_inserted} $+$ \texttt{\#lines\_deleted} \cite{fan2018early}\\
     &\texttt{\#files}&Number of files changed in the first revision \cite{fan2018early}\\
     &\texttt{\#files\_type}&Number of file extension in the change \cite{fan2018early}\\
     &\texttt{\#directory}&Number of change directory \cite{fan2018early}\\
     &\texttt{\#segs\_added}&Number of code segments added \cite{fan2018early}\\
     &\texttt{\#segs\_deleted}&Number of code segments deleted \cite{fan2018early}\\
     &\texttt{\#segs\_modify} &Number of code segments modified \cite{fan2018early}\\
      &\texttt{change\_entropy} & Code change modification distribution across the changed files \cite{fan2018early}\\
     \hline
    \multirow{7}{*}{Text metrics}&\texttt{subject\_length} &Number of characters in the suject of the changes\\
     &\texttt{subject\_word\_count}&Number of characters in the subject of the code change\\
     &\texttt{msg\_length} &Number of characters in the description of code change\\
     &\texttt{msg\_word\_count}&Number of words in the description of code change \cite{fan2018early}\\
     &\texttt{is\_non\_fonctional} &True if the change description contains the keywords related to non-functional additions\\
     &\texttt{is\_perfective}&True if the change description contains the keywords related to code refinement\\
     &\texttt{is\_refactoring} &True if the change description contains the keywords related to refactoring \cite{fan2018early}\\
     \hline
    \multirow{12}{*}{\begin{tabular}[C]{l} Owner experience \\ metrics\end{tabular}}  &\texttt{\#owner\_prior\_changes}&The number of prior changes submitted by the owner \cite{fan2018early}.\\
     &\texttt{\#prior\_merged\_changes} &The number of prior merged code reviews by the owner. \\
     &\texttt{\#prior\_abandoned\_changes}&The number of prior abandoned code reviews by the owner.\\
     &\texttt{merge\_ratio} &ratio of merged changes. \cite{fan2018early}\\
     &\texttt{\#prior\_subsystem\_changes}&Subsystem prior code reviews count \cite{fan2018early}.\\
     &\texttt{prior\_code\_reviews\_duration (min, max, avg, std)}&The completion time of the previous code reviews of the owner (min,max,avg,std).\\
    
     &\texttt{\#prior\_owner\_subsystem\_changes} &The number of prior subsystems changed done by the owner.\\
     &\texttt{prior\_owner\_subsystem\_changes\_ratio} &Ratio  of prior subsystem changes count by the number of prior changes done by the owner.\\
     &\texttt{\#reviewed\_changes\_owner} &The number of the reviewed code reviews by the owner.\\
     &\texttt{\#owner\_previous\_message} &The number of messages sent by the owner in previous changes.\\
     &\texttt{\#owner\_exchanged\_messages} &Total number of exchanged messages in the owner's changes.\\
     &\texttt{\#owner\_messages\_avg\_per\_changes (min, max, avg, std)} &The number of exchanged messages in the owner's changes per change (min, max, avg, std).\\
     \hline
    \multirow{3}{*}{File history metrics} &\texttt{files\_changes\_duration (min, max, avg, std)} &Completion time of code reviews involving at least one file of the current code review files.\\
     &\texttt{\#developers\_file} &The number of developers of the code review files \cite{fan2018early}.\\
     &\texttt{\#prior\_changes\_files} &The number of code reviews involving at least one of the current code review files count \cite{fan2018early}.\\
    \hline
     \end{tabular}
     
%     }
     \label{tab:features}
\end{table*}

\subsection{Build and validate regression models}
In this step, we build and validate regression models to predict code review completion time (in hours). In this study, our outcome variable is the difference is $T_{c} - T_{s}$ where $T_{c}$ denotes the time when the code review is done (\ie either merged or abandoned ) and $T_{s}$ denotes the moment of the creation the given code review. We use the training data extracted in previous steps to build regression models and validate their performances. We will build and validate our ML classifiers using \textit{Scikit learn}\footnote{\url{https://scikit-learn.org/stable/index.html}}. To build our classifiers, we follow a three-step process as summarized in Table \ref{tab:ml_steps}. In particular, we proceed as follows:
\begin{table*}[ht!]
	\fontsize{7}{8.5}\selectfont
	\tabcolsep=0.15cm
     \center
     \caption{ML steps configurations.}
%     \resizebox{\textwidth}{!}{
     \begin{tabular}{p{2.2cm}p{3.5cm}p{7.5cm}}
     \hline
     \textbf{Step}&\textbf{Options} & \textbf{Scikit learn algorithm}  \\
     
     \hline
     \multirow{2}{3.5cm}{(1) Feature selection}&Recursive feature selection& \texttt{sklearn.feature\_selection.RFECV}\\&Sequential feature selection &\texttt{sklearn.feature\_selection.SequentialFeatureSelector}\\
     \hline
     \multirow{3}{3.5cm}{(2) Data normalization}&None& \\&Min-max normalization&\texttt{sklearn.preprocessing.MinMaxScaler}\\&
     Standardization&\texttt{sklearn.preprocessing.StandardScaler}\\
     \hline
     \multirow{11}{3.5cm}{(3) ML regressors}&Linear regression (LR)&\texttt{sklearn.linear\_model.LinearRegression}
     \\&Lasso Regression (LaR)& \texttt{sklearn.linear\_model.Lasso}
     \\&Ridge Regression (RR)&\texttt{sklearn.linear\_model.Ridge}
     \\&Bayesian Lasso Regression (BLaR)&\texttt{sklearn.linear\_model.BayesianRidge}
     \\&Support Vector Machine (SVM)&\texttt{sklearn.svm.SVR}
     \\&k-Nearest Neighbor (kNN)&\texttt{sklearn.neighbors.KNeighborsRegressor}
     \\&Decision Tree (DT)&\texttt{sklearn.tree.DecisionTreeRegressor}
     \\&Neural Networks (NN)&\texttt{sklearn.neural\_network.MLPRegressor}
     \\&Random Forrest (RF)&\texttt{sklearn.ensemble.RandomForestRegressor}
     \\&AdaBoost with DT (AdaDT)&\texttt{sklearn.ensemble.AdaBoostRegressor}
     \\&Gradient Boost (GB)&\texttt{sklearn.ensemble.GradientBoostingRegressor}
     \\
     \hline
     
     \end{tabular}
     
%     }
     \label{tab:ml_steps}
\end{table*}
\paragraph{Feature selection} Having a high number of features can lead to noise in the dataset since features may hold redundant information and make the model training slower. Since the code review completion time has a high number of features, we plan to test two different feature selection techniques, namely the recursive and sequential feature selection, as presented in Table \ref{tab:ml_steps}.

\paragraph{Data normalization}
ML classifiers typically require features to have a close scale \cite{singh2020investigating}. The scale difference between features can influence the performance of an ML regressor. Hence, we test different data normalization scenarios, namely min-max scaling and standardization as compared to the original data without normalization (cf. Table \ref{tab:ml_steps}).

\paragraph{ML regressors fitting}
\label{sec:ml_regressors}
The main goal of our framework is to study the effectiveness of predicting code review duration using ML algorithms. Hence, we test the performance of multiple ML regressors namely, Linear regression (LR), Lasso Regression (LaR), Bayesian Lasso Regression (BLaR), Ridge Regression (RR), Support Vector Machine (SVM), k-Nearest Neighbor (kNN), Decision Tree (DT), Random Forrest (RF), AdaBoost with DT (AdaDT), Gradient Boost (GB) and Neural Networks (NN). Note that our fitting ML regressors involve learning their hyper-parameters since their values have proven to affect ML algorithms' performances in several software engineering prediction problems \cite{tantithamthavorn2016automated}. To this end, we plan to apply grid search to learn the best hyper-parameters for each algorithm to ensure a fair comparison. 
\subsection{Apply the built model to new data} 
After training and validating the performance of each ML regressor, we select the best regressor model based on evaluation metrics defined in \ref{sec:evaluation_metrics}. Finally, we apply the selected model to a given code review to predict its review completion time.

\subsection{Rank features} 
\label{sec:rank_feature}
One of the goals of our framework is to understand the most influential features impacting the delay of code review. Knowing this piece of information can help developers and project managers to plan informed retro-actions leading to predict and reduce review delays. To measure the importance of the features, we use the Leave-One-Covariate-Out (LOCO) approach proposed by Lei et al. \cite{lei2018distribution} since it is a model-free approach and focuses directly on predictive quantities. 

\section{Empirical study design}
\label{sec:empirical_study}
In this section, we start by setting the research questions. Later, we describe the experimental process that will drive our study in order to answer our research questions. Finally, we present the details related to our planned experimental study to validate our proposal.

\subsection{Research questions} 
The main goal of our approach is to investigate whether it is possible to learn and predict the completion time of a given code review. In addition, we plan to understand the main factor that impacts the delay of a code review. Our research is driven by the following research questions: 
\begin{enumerate}

\item[\textbf{\textit{RQ1.}}] \textbf{(ML algorithms performance):} \textit{How effective are machine learning models for code review duration prediction?}

\item[\textbf{\textit{RQ2.}}]
\textbf{(All features vs features subsets):} \textit{How effective is our prediction model when all features are used than when only a subset is used?} %\ali{not sure how relevant is this RQ? The features importance will be studies in RQ3.}
%\Moataz{this research question is similar to RQ2 here: \url{https://link.springer.com/content/pdf/10.1007/s10664-018-9602-0.pdf}}

\item[\textbf{\textit{RQ3.}}] \textbf{(Features importance analysis):} \textit{What are the most important features influencing code review completion time?}
\end{enumerate}

\subsection{Experimental setting overview}
\label{sec:exp_setting}
\begin{figure}[!h]
    \centering
    \includegraphics[width=\columnwidth]{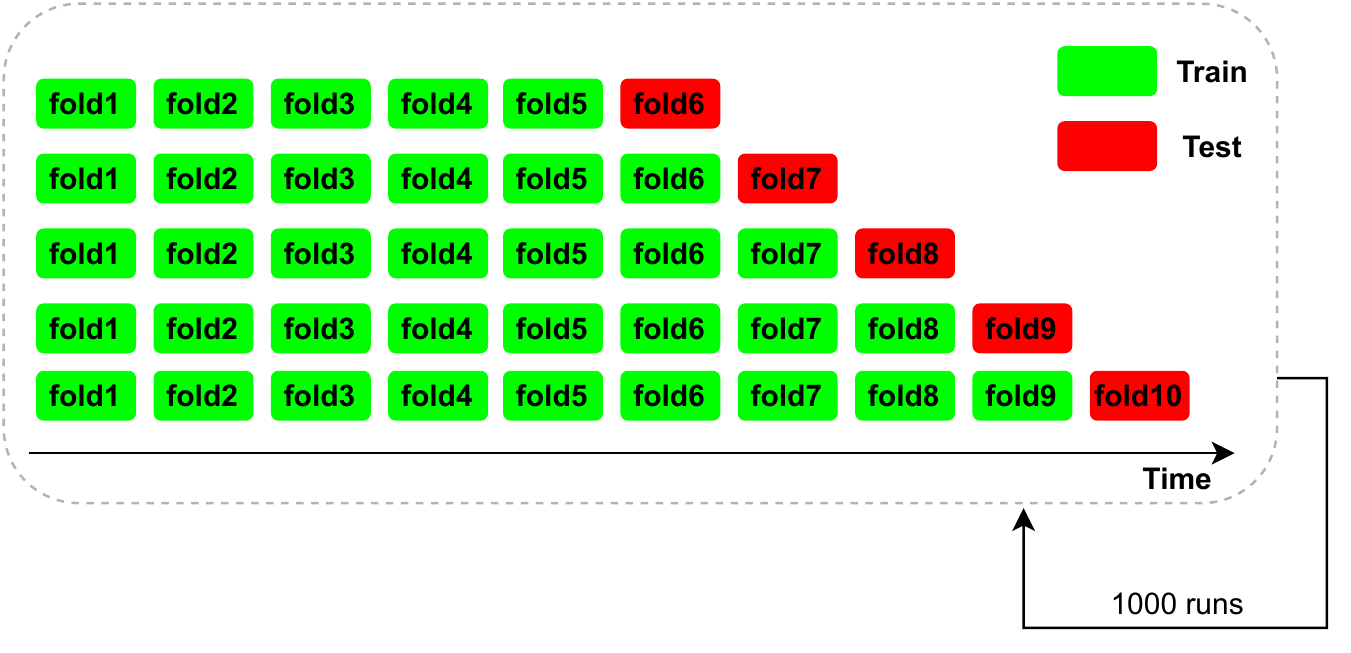}
    \caption{Online validation process.}
    \label{fig:validation_process}
\end{figure}

In practice, code review data changes over time, especially for long-lived projects. Thus, it is important to consider the code review completion time prediction as an online learning problem since it is more similar to what happens in practice \cite{tan2015online,saidani2020predicting}.  

To measure the performance of the different ML algorithms, we use the online validation process depicted in Figure \ref{fig:validation_process}. Initially, we sort code reviews according to their creation time in ascending order. Then, we split the whole dataset into 10  folds of equal sizes. Thereafter, we select the last 5 folds as testing data. Then, for each iteration $i$ where $1 \leq  i \leq 5$, we use the $fold_{i+5}$ as a testing set, while we keep the earlier folds $fold_{1,2,..,i+4}$ as training set. This process is repeated 1000 times to alleviate the stochastic nature of the data as well as the ML algorithms (except DT and KNN since it is deterministic).

\subsection{Data collection}
\label{sec:datasets}
We plan to validate the results of our study using three OSS projects, namely, LibreOffice\footnote{https://gerrit.libreoffice.org}, Openstack\footnote{https://review.opendev.org/} and QT\footnote{https://codereview.qt-project.org/} projects data since they have been widely studied in  previous research in MCR  \cite{fan2018early,thongtanunam2020review,ruangwan2019impact,ouni2016search,chouchen2021whoreview}. We present in Table \ref{tab:studied_projects} a summary of our  collected data that we plan to use in our experiments. 

\begin{table}[ht]
    \centering
     \caption{Studied projects.}
    \begin{tabular}{cccc}
    \hline 
    \textbf{Project}&\textbf{Period}&\textbf{\#Code changes}&\textbf{\#developers}  \\
    \hline
    LibreOffice&06.03.2012 - 27.04.2021&110515& 1,219\\
    Openstack&16.12.2012 -  27.04.2021&784011&15,432\\ 
    QT&17.05.2011- 27.04.2021&293170& 3,264\\
    
    \hline 
    \end{tabular}
   
    \label{tab:studied_projects}
\end{table}

\subsection{Performance metrics}
\label{sec:evaluation_metrics}
To evaluate our different ML algorithms we refer to three standard regression metrics: Mean Absolute Error (MAE) and Mean Relative Error (MRE) since they have been used in a similar context \cite{maddila2019predicting}. Additionally, we also use Standard Accuracy (SA) since it was widely used in evaluating effort estimation approaches in software engineering \cite{tawosi2021multi}.

\section{Analysis method}
\label{sec:analysis_method}
In this section, we present the analysis method for each RQ. 
\subsection{Analysis method for RQ1}
The main goals of RQ1 are to see (\textit{i}) whether it is possible to use ML regressors to predict the code review completion time, and (\textit{ii}) identify the best ML regressor for the task. 
To answer RQ1, we evaluate all ML regressors presented in Section \ref{sec:ml_regressors}. To compare the performance of the different algorithms, we follow the experimental process established in 
Section \ref{sec:exp_setting}using our three performance metrics MAE, MRE, and SA (cf. Section \ref{sec:evaluation_metrics}). 

Thereafter, we report the results and discuss the performance achieved by each ML regressor for each dataset using mean and median values.% along with Wilcoxon test and Cliff's d statistic defined in \ref{sec:statistical_tests}.
To verify the performance difference statistically for different experiments in this RQ, we will apply a Wilcoxon signed rank test \cite{wilcoxon1992individual} at a level of confidence 99\% with a Bonferroni correction \cite{weisstein2004bonferroni} to alleviate the effect of multiple comparisons. In addition, we apply the Cliff's delta effect size test \cite{romano2006appropriate}. The magnitude $|d|$ as the effect size is interpreted as follows:  
negligible (N) if $|d|<0.147$, small (S) if $0.147\leq|d|<0.33$, medium (M) if $0.33\leq|d|<0.474$) and large (L) $|d|\geq0.474$. Specifically, we use the implementations of both tests in R. 

\subsection{Analysis method for RQ2}
In this study, we used 50 features divided into six main dimensions as shown in Table \ref{tab:features}. In this RQ, we aim to study the performance of ML regressors when using all dimensions of data compared to using only one dimension to see the benefit of exploiting knowledge from all dimensions together. Once identified in RQ1, we use the best ML regressor and study its performance when using all features dimensions against using only one dimension at once by applying the validation process described in Section \ref{sec:exp_setting}. Then, we report the results of the comparisons for each dataset based on performance metrics and statistical tests. Having better performance when using all features dimensions indicate the usefulness of using all dimensions together.

\subsection{Analysis method for RQ3}
Identifying the main factors impacting code review completion time is crucial to help developers improve the code review process. The main goal of RQ3 is to rank accurately the features presented in Table \ref{tab:features} according to their importance for each experimental dataset (cf. Section \ref{sec:datasets}). To measure the importance of each metric, we use the LOCO method introduced in \ref{sec:rank_feature}. We will follow our validation process defined in \ref{sec:exp_setting} to run multiple experiments for each feature. %Finally, we will assess all the obtained results under the ESD test to rank the features according to their importance. 

To rank the feature importance, we will use the Scott-Knott Effect Size Difference (ESD) test elaborated by Tantithamthavorn et al. \cite{tantithamthavorn2018optimization}. ESD is similar to Scott-Knott
test with slight two changes: 
\begin{enumerate}
    \item Normality correction: Tantithamthavorn et al. \cite{tantithamthavorn2018optimization} applied log-transformation $y = log(x+1)$ to alleviate the skewness of the data. \item Effect size correction: Tantithamthavorn et al. used Cohen's $d$ \cite{cohen2013statistical} to measure the effect size between different clusters and merged clusters having negligible effect size, \ie, having $d < 0.2$. 
\end{enumerate}

For ESD, we use the implementation provided by Tantithamthavorn et al.\footnote{https://github.com/klainfo/ScottKnottESD/tree/master} since it was updated according to the recommendation of Herbold \cite{herbold2017comments} allowing us to verify Scott-Knott assumptions prior to apply the test.

\section{Publication of the experimental dataset}
To allow replicating and extending our study, the data collected data from our study will be publicly available
using an online open-access repository (\eg, Github). We plan to release the collected data, both dependent and independent variables, from each studied project.

\balance

\bibliography{references}
\bibliographystyle{IEEEtran}

\end{document}